%% file: main.tex
\documentclass[10pt,sigconf,letterpaper,nonacm]{acmart}

\usepackage{tikz}
\usepackage{amsmath}

\usepackage{pifont,csquotes}
\usepackage[official]{eurosym}
\usepackage[multiple]{footmisc}
\usepackage{epsfig,endnotes}
\PassOptionsToPackage{hyphens}{url}
\usepackage{url}
\usepackage{color}
\usepackage{subfigure}
\usepackage{colortbl}
\usepackage{multirow}
\usepackage{xcolor}
\usepackage{comment}
\usepackage{textcomp}
\usepackage[normalem]{ulem}
\usepackage{xspace}
\usepackage{soul}
\usepackage{textcomp}

\usepackage[T1, OT1]{fontenc}

\usepackage{calc}

\usepackage{enumitem}

\DeclareTextSymbolDefault{\dh}{T1}

\newenvironment{packed_itemize}{
\begin{itemize}[leftmargin=*]
  \setlength{\itemsep}{4pt}
  \setlength{\parskip}{0pt}
  \setlength{\parsep}{0pt}
  \setlength{\topsep}{2pt}
}{\end{itemize}}

\newenvironment{packed_enumerate}{
	\begin{enumerate}[leftmargin=*,label=(\alph*)]
  \setlength{\itemsep}{2pt}
  \setlength{\parskip}{0pt}
  \setlength{\parsep}{0pt}
  \setlength{\topsep}{2pt}
}{\end{enumerate}}

\newcommand{\deltwo}[1]{}

\newcommand{\del}[1]{}

\newcommand{\eg}{e.g.,\ }

\newcommand{\roa}{\textsf{ROA}\xspace}
\newcommand{\roas}{\textsf{ROAs}\xspace}

\newcommand{\aspas}{\textsf{ASPAs}\xspace}

\newcommand{\vrps}{\textsf{VRPs}\xspace}

\makeatletter
\renewcommand{\sectionautorefname}{\S\@gobble}
\renewcommand{\subsectionautorefname}{\S\@gobble}
\renewcommand{\subsubsectionautorefname}{\S\@gobble}
\makeatother

\newcommand{\ourparagraph}[1]{\par\medskip\noindent\textbf{#1}\hspace{0.1em}}

\newcommand{\ours}{\textsf{pqRPKI}\xspace} % Improve.

\usepackage[draft,inline,nomargin,index]{fixme}
\fxsetup{theme=colorsig,mode=multiuser,inlineface=\itshape,envface=\itshape}

\title{pqRPKI: A Practical RPKI Architecture for the Post-Quantum Era}

\input{abstract}

\author{Weitong Li}
\affiliation{%
  \institution{Virginia Tech}
  \city{Blacksburg}
  \state{VA}
  \country{USA}
}

\author{Yuze Li}
\affiliation{%
  \institution{Virginia Tech}
  \city{Blacksburg}
  \state{VA}
  \country{USA}
}

\author{Taejoong Chung}
\affiliation{%
  \institution{Virginia Tech}
  \city{Blacksburg}
  \state{VA}
  \country{USA}
}

\begin{document}
\maketitle

\input{introduction}
\input{background}

\input{design}
\input{evaluation}

\input{discussion}
\input{conclusion}

\input{paper.bbl}

\end{document}

%% file: abstract.tex
\begin{abstract}
The Resource Public Key Infrastructure (RPKI) secures Internet routing by binding IP prefixes to authorized Autonomous Systems, yet its RSA foundations are vulnerable to quantum adversaries.
A na\"ive swap to post-quantum (PQ) signatures (\eg Falcon) is a poor fit for RPKI's bulk model: every relying party (RP) repeatedly fetches and validates the entire global repository, so larger keys and signatures inflate bandwidth and CPU cost, especially during a long dual-stack transition.

We present \ours, a post-quantum RPKI framework that pairs a multi-layer Merkle Tree Ladder (MTL) with RPKI objects, customized to relocate per-object verification material from certificates into the Manifest.
To update RPKI for Merkle tree based schemes, \ours redesign the RPKI manifest and delegation chain, introduces a ladder-guided sync and bulk-verification workflow that lets validators localize diffs top-down and rebuild trees bottom-up.

\ours also preserves current RPKI objects and encodings, supports both hosted and delegated operation, and provides an additive migration path that coexists with today's trust anchors for dual-stack deployment with little size overhead.
Implemented as a working publication point (PP) and RPs, we show that pqRPKI reduces repository footprint to 546.8 MB on average (65.5\%/83.1\% smaller than Falcon/ML-DSA), cuts full-cycle validation to 102.7 s, and achieves 118.3 s end-to-end PP to Router time, enabling sub-2-minute operating cadences with full-repository validation each cycle. Dual-stack deployment with RSA only adds just 3.4\% size overhead versus today's RPKI repositories.
\end{abstract}

%% file: introduction.tex
\section{Introduction}
\label{sec:introduction}
The Resource Public Key Infrastructure (RPKI)~\cite{rfc6480} is the Internet's deployed mechanism for authenticating interdomain routing intent.
It binds Internet number resources (IP prefixes and ASNs) to the entities authorized to originate them, and lets operators publish signed routing policy objects---most notably Route Origin Authorizations (\roas)~\cite{rfc9582}.
Relying parties (RPs) fetch repository contents from publication points, validate certificate and object chains, and output validated \roa payloads (VRPs) that routers use to reject invalid route origins.
Unlike handshake-driven PKIs, this pipeline is inherently \textit{bulk}: each RP periodically retrieves and validates (essentially) the complete global corpus across all trust anchors and delegated publication points, on an operational cadence (commonly 20$\sim$60 minutes, and at least every 24 hours).

This bulk model is exactly what makes the post-quantum transition urgent \textit{and} difficult; RPKI's security today rests on classical signatures (RSA-2048), which will not withstand large-scale quantum adversaries.
A straightforward response is to replace RSA with NIST-standard post-quantum signatures such as Falcon or ML-DSA/Dilithium~\cite{kwiatkowski-2019-pqc}.
However, na\"ively swapping the signature algorithm is a poor fit for RPKI at Internet scale: PQ public keys and signatures are substantially larger, verification is costlier, and these costs are multiplied across \textit{hundreds of thousands of repository objects} that \textit{every RP must download and validate every cycle}.
Worse, a practical migration must support a long dual-stack period for backward compatibility~\cite{GooglePQCThreatModel}---meaning repositories may need to present \textit{both} classical and post-quantum authentication material, compounding storage, bandwidth, and synchronization overhead for publication points and validators.

Crucially, RPKI is not just a victim of this scaling problem; it also has structure we can exploit.
Compared to WebPKI~\cite{valenta-2025-webpki}, RPKI validators \textit{must} fetch complete repositories on a cadence, and each validation cycle already depends on two mandatory, frequently refreshed artifacts: the Manifest and the CRL.
Validation is also naturally \textit{bulk} under each CA: an RP typically verifies many objects issued by the same CA in one run.
These properties create a repository-native metadata channel (the Manifest) and repeated batch-processing opportunities that do not exist in handshake-bounded PKIs.
Prior post-quantum RPKI explorations that simply dropped PQ signatures into today's per-object certificate layout did not leverage this environment, and therefore inherited worst-case overhead~\cite{doesburg-2025-post}.

These observations motivate a different approach: instead of forcing post-quantum signatures into self-contained, per-object certificates, we reorganize where authentication material lives and how validators use it so that RPKI's \textit{existing} bulk workflow carries more of the load---while keeping current RPKI objects and structure intact for backward compatibility and dual-stack operation.

We present \ours, a post-quantum RPKI architecture that exploits RPKI's repository model.
\ours combines the RPKI Manifest with a Merkle Tree Ladder (MTL)~\cite{harvey-2023-mtl} per CA, and makes RPKI-specific design moves that preserve operational deployability:
(i) we relocate per-object Merkle verification material (leaf index and authentication information) \textit{out of per-object certificates and into the Manifest}, avoiding per-object bloat and update churn;
(ii) we use a multi-level MTL organization that supports both hosted and delegated CA operation with minimal additional overhead;
(iii) we model the frequently reissued Manifest and CRL as \textit{depth-0 rungs} appended after object rungs, so routine metadata refreshes do not perturb object leaves; and
(iv) we drive synchronization and validation from the authenticated Manifest, enabling ladder-guided change localization and bottom-up bulk verification.

\ours preserves today's \roa objects, supports both hosted and delegated CA operation, and extends the Manifest \textit{additively} so legacy validators can continue operating unchanged during a long dual-stack migration.
Our main contributions are:
\begin{packed_itemize}
  \item \textit{An RPKI-tailored post-quantum authentication architecture.}
  We introduce pqRPKI, which combines a multi-level Merkle Tree Ladder with the existing RPKI repository structure, and redesigns MTL for RPKI to keep objects small and stable under churn.

  \item \textit{A manifest-driven synchronization and bulk-verification workflow.}
  We design a ladder-guided RP workflow that localizes diffs at the authenticated Manifest and reconstructs Merkle structure bottom-up, reducing bytes in flight and avoiding redundant per-object work.

  \item \textit{Operational compatibility and a practical migration path.}
  pqRPKI keeps \roa objects unchanged, preserves RSA for dual-stack deployment, supports both hosted and delegated modes, and requires only additive extensions so incremental deployment remains feasible.

  \item \textit{A working prototype and evaluation on production repositories.}
  We implement a pqRPKI publication point and relying party and evaluate over seven days of production repositories.
  pqRPKI reduces average repository size to 546.8~MB in PQC-only mode (65.5\%/83.1\% smaller than na\"ive FN-DSA/ML-DSA), cuts full-cycle validation to 102.7~s, and supports sub-2-minute operation with full-repository validation each cycle.
  In dual-stack mode, pqRPKI adds only 3.4\% repository-size overhead over today's RSA-based RPKI.
\end{packed_itemize}

%% file: background.tex
\section{Background and Motivation} \label{sec:background}
\subsection{RPKI Overview}
The Border Gateway Protocol (BGP)~\cite{rfc4271} is the Internet's interdomain routing protocol: Autonomous Systems (ASes) exchange reachability for IP prefixes.
BGP lacks origin authentication, enabling prefix hijacks that cause traffic interception and service disruption~\cite{butler-2010-bgp, cowie-2010-china, ballani-2007-ipprefix}.
The Resource Public Key Infrastructure (RPKI)~\cite{rfc6480} addresses this authorization gap.
RPKI binds IP prefixes and ASNs to cryptographic keys via a hierarchical, X.509-based PKI rooted at the five Regional Internet Registries (RIRs).
Each RIR operates a trust-anchor certificate attesting to its authority over a set of number resources; certificates then delegate subsets down to National Internet Registries (NIRs), ISPs, and organizations.

\subsubsection{RPKI Structures}
RPKI consists of resource certificates and signed objects arranged in a delegation hierarchy.
Each RIR is a trust anchor with a self-signed certificate; it issues subordinate resource certificates to allocate resources, forming certificate chains.
Each Certificate Authority (CA) maintains a key pair and publishes its signed artifacts in a repository known as a \textit{publication point} (PP).

Local Internet Registries (LIRs) typically use one of two operational modes.
In \textit{delegated RPKI}, an LIR runs its own CA and repository, retaining control over keys and publication.
Because operating a CA and PP is operationally burdensome, many LIRs instead use \textit{hosted RPKI}, where the RIR operates the LIR's CA: the RIR issues a distinct CA certificate for the LIR and manages and publishes all signed objects, while the LIR interacts via a portal~\cite{ARINHostedRPKI}.
Hosted LIRs remain logically distinct CAs, but day-to-day CA operations reside with the RIR.

\subsubsection{RPKI Objects}
RPKI defines several signed object types, encoded in ASN.1~\cite{rfc5911} and encapsulated in CMS~\cite{rfc5652}; the standardized signature algorithm is RSA-2048 with SHA-256~\cite{rfc8017, rfc5754}:
\begin{packed_itemize}
    \item \textit{CA certificates} (resource certificates) bind a public key to specific IP prefixes and/or AS numbers.
    \item \textit{EE certificates} are \textit{one-use certificates}: each EE key signs exactly one RPKI signed object (the RFC ``one EE per object'' model~\cite{rfc6487}). EE certificates are commonly embedded within the \roa object they sign.
    \item \textit{RPKI signed objects} encode routing policy artifacts---most notably \textit{Route Origin Authorizations (\roas)}~\cite{rfc9582}, which authorize an AS to originate a prefix. RPKI Signed Checklists (RSCs)~\cite{rfc9323} are another signed object type used for non-public artifacts; although not published to public repositories, they still rely on EE certificates for validation and revocation.
    \item \textit{Manifests} enumerate, per CA, the complete file set and hashes, allowing relying parties to detect missing, stale, or substituted objects.
    \item \textit{Certificate Revocation Lists (CRLs)} are X.509 structures listing revoked certificates, checked during validation.
\end{packed_itemize}

As of August 13, 2025 (our measurement snapshot), the public RPKI ecosystem contains 64 repositories and about 465,932 objects: 47,739 CA certificates, 49,263 Manifests, 49,262 CRLs, and 319,186 \roas (plus experimental sets of other object types).
The total repository footprint is about 886.6~MB, and the five RIR repositories serve most objects (87.8\%).
RPKI also has clear headroom to grow: \roa coverage remains roughly 58.8\%~\cite{NISTRPKIMonitor}, and many operators have not yet created \roas for all prefixes.
Both trends imply larger repositories and longer validation cycles over time.
%Meanwhile, new object types (\eg ASPA~\cite{ASPAProfile}) are being introduced.

\subsubsection{RPKI Publication}
Publication points host all CA-issued artifacts: CA and EE certificates, signed objects (\eg \roas), Manifests, and CRLs.
A distinctive operational property of RPKI is intentionally short-lived issuance.
Manifests and their corresponding EE certificates commonly have validity periods of about 8-24 hours (\eg Krill defaults)~\cite{Krill}, shrinking the window in which stale or compromised artifacts can be abused.
On expiry, CAs regenerate EE key pairs and re-sign the corresponding objects, issuing a new Manifest reflecting repository contents.
As a result, EE certificates and Manifests are frequently re-signed even when the underlying routing intent is unchanged.
A recent measurement~\cite{madory-2024-timesup} reports RPKI object validity periods ranging from roughly 8~hours (ARIN) to 3~days (LACNIC) in current deployments.

\subsubsection{RPKI Synchronization and Validation}
Network operators run Relying Parties (RPs) that periodically synchronize and validate RPKI.
Starting from trust anchors, an RP fetches the corresponding certificates and follows subordinate chains to discover and retrieve \textit{all reachable repositories}.

Repositories are typically accessed via the RPKI Repository Delta Protocol (RRDP)~\cite{rfc8182}, which replaced early \texttt{rsync} deployments due to security and scalability issues.
RRDP runs over HTTPS and uses XML; a versioned \textit{notification file} advertises a session identifier, the current snapshot serial, and links to delta files for prior serials.
Deltas encode additions, deletions, and replacements between serials.
On startup, an RP fetches the notification file and latest snapshot; if its cached serial appears in the notification, it applies the listed deltas.
Otherwise (\eg after restarts or missed updates), the RP refetches the full snapshot to resynchronize even if most object files are unchanged, which can impose substantial bandwidth and delay on large repositories.

After synchronization, the RP validates signatures, verifies certificate chains, and checks CRLs, then emits validated \vrps for routers to filter invalid routes.
Recent measurements~\cite{sediqi-2025-rpki} report about 490~s to download and validate from an empty cache, and about 210~s on average for continuous RRDP updates at a 20-minute interval.

\subsection{Post-Quantum Cryptography: A Brief Primer}
Post-quantum cryptography (PQC) resists quantum adversaries using lattice, hash, code, and multivariate primitives.
NIST has selected one lattice-based KEM (ML-KEM) and three digital signatures: the lattice-based ML-DSA and Falcon, and the hash-based SLH-DSA~\cite{nist-ML-DSA, fouque-2020-falcon}.
These schemes trade off key/signature size and computation, but PQ signatures are typically much larger than today's Internet-deployed RSA.
Even Falcon, the most compact option, uses 666~B signatures and 897~B keys---$\sim$3$\times$ larger than RPKI's RSA-2048.

Because large-scale quantum computers are not yet available and PQC introduces real bandwidth/CPU overheads, deployments are expected to run a long-lived dual-stack transition (classical + PQ) to preserve backward compatibility~\cite{GooglePQCThreatModel}.
Accordingly, a PQ-ready Internet PKI must account for both cryptographic migration and operational budgets.

A straightforward approach is to swap existing signatures for PQ variants, or deploy both in parallel (\eg hybrid signatures);
in WebPKI/TLS, hybrids such as ECDSA+Dilithium have been explored to maintain compatibility while preparing for PQ threats.
Because TLS handshakes carry only a small number of certificates and signatures, larger PQ material is often manageable, though it can still affect latency and fragmentation.

\ourparagraph{Why a na\"ive PQ swap strains RPKI.}
RPKI is unusually sensitive to PQ overheads because:
\begin{packed_itemize}
    \item \textit{Object volume and composition.} Repositories contain large numbers of certificates (especially one-use EE certs) and signed objects (\eg \roas).
    \item \textit{Signing churn.} Many objects are reissued frequently, particularly Manifests (hours to $\sim$1 day) and their corresponding EE certificates.
    \item \textit{Global, frequent validation.} RPs download and validate \textit{all} objects from \textit{all} repositories on a tight cadence to keep routers up to date.
\end{packed_itemize}

A recent study~\cite{ddoesburg-2025-pqc} found that adopting NIST-standard multi-use PQ signatures in today's RPKI organization substantially increases repository size and RP validation time; even ``size-saving'' tweaks remain well above current RSA-based deployments.
Across Falcon/Dilithium-class schemes, the total repository footprint would grow by roughly 1.7-4.0$\times$ and RP validation time by about 1.8-2.6$\times$, straining bandwidth and processing budgets.
At ecosystem scale (thousands of RPs) and under a necessary dual-stack transition, the aggregate load on PPs and RPs only increases.

Taken together, these results suggest that simply swapping RSA for a standardized multi-use PQ signature is unlikely to be practical at full RPKI scale.

\subsection{Compressing RPKI Repositories}
The growth of RPKI repositories is an operational concern even without PQC.
Two recent proposals, the ``null scheme''~\cite{ddoesburg-2025-pqc} and iRPKI~\cite{schulmann-2025-pruning}, reduce repository size by removing parts of today's object set (e.g., EE certificates, and in some cases even \roas) and by applying size-reduction techniques such as more compact encodings than X.509/CMS.

These proposals can substantially reduce the number of certificates and total bytes, and could therefore offset some PQ signature overhead.
However, the IETF Sidrops working group has raised concerns that removing EE certificates and other object content breaks existing RPKI structures and workflows~\cite{SidropsNonViableProposal2025,SidropsPQCforRPKI2025}.
In addition, backward compatibility and a long dual-stack transition (classical + PQ) further complicate adoption, similar to PQ transitions in other PKI ecosystems~\cite{doesburg-2025-post}.

While the debate over how much ``integrity'' RPKI repositories must preserve is ongoing in the IETF, our goal is to explore a new authentication structure that can satisfy the integrity and backward-compatibility requirements raised in the working group, while also being PQ-ready.
At the same time, our structure is compatible with byte-reduction techniques like iRPKI and the null scheme when deployments choose to apply them.

\subsection{Why Existing Merkle and MTL Designs Do Not Fit RPKI}
An alternative to pruning certificates is a Merkle tree-based authentication: hash many objects into a tree and protect only the root with a post-quantum signature.
This amortizes expensive PQ signatures across many objects and can drastically reduce published bytes.

Merkle designs, however, introduce practical challenges.
Validation requires not only the root signature but also {\em per-object authentication paths}, thus object updates can change the root and intermediate nodes, potentially forcing frequent re-signing.
Prior work has addressed these issues mainly in DNSSEC~\cite{hoang-2025-pqdnssec}, most notably via Merkle Tree Ladder (MTL) mode~\cite{harvey-2023-mtl}, which composes many bounded-depth trees (``rungs'') to limit re-signing and embeds authentication paths with individual records.

However, directly applying these designs to RPKI is not straightforward.
First, RPKI repositories contain hundreds of thousands of objects; embedding per-object authentication paths would add tens of hashes per object (\eg $\sim$640~B with SHA-256), substantially inflating repository size.
Second, RPKI has a multi-level delegation hierarchy (RIRs to LIRs) and supports both hosted and delegated operation, which DNSSEC-oriented designs do not model.
Third, RPKI validation is mandatory and bulk (entire repositories each cycle), whereas existing Merkle designs largely optimize for standalone, per-object validation.

%% file: design.tex
\section{Design of \ours}
\label{sec:design}

\ourparagraph{Design Goals}
Our goal is a post-quantum-secure RPKI that remains operationally viable at Internet scale and throughout a long transition period. We target:
\begin{packed_itemize}
    \item \textit{Post-quantum security.}
    Authenticate repository state with NIST-approved post-quantum signature schemes.

    \item \textit{Practical size, signing, and validation.}
    Keep published bytes and RP work small enough for frequent synchronization and full-cycle validation, while still supporting per-object checks where required (\eg non-public RSC workflows).

    \item \textit{Dynamic updates and revocation.}
    Support frequent reissuance, revocation, and incremental changes without broad re-signing or churn across unrelated objects.

    \item \textit{Incremental deployment and compatibility.}
    Support long-lived dual deployment (RSA + PQ) without changing existing RPKI payloads or encodings (\eg \roas).
    In particular, \ours extends the Manifest additively so legacy validators can continue validating RSA unchanged while upgraded validators validate \ours in parallel.
\end{packed_itemize}

\ourparagraph{Deployability constraint.}
Several proposals reduce repository size by restructuring RPKI objects, for example by removing EE certificates~\cite{ddoesburg-2025-pqc} or pruning fields and changing encodings/semantics~\cite{schulmann-2025-pruning}.
These directions can yield substantial byte savings, but they also move away from today's object model and workflows, which has been raised as a concern in Sidrops~\cite{SidropsNonViableProposal2025,SidropsPQCforRPKI2025}, and they can complicate a long transition that requires backward compatibility.
Accordingly, \ours is minimally invasive: it preserves existing RPKI payloads and hierarchy and layers PQ authentication alongside RSA, enabling incremental deployment without immediate changes to validators or CA toolchains.

\ourparagraph{Design intuition: treat the Manifest as the shared verification channel.}
RPKI has properties that make a repository-first design practical, and that do not hold for handshake-driven PKIs:
\begin{packed_itemize}
    \item \textit{Global, mandatory bulk fetch.}
    Unlike TLS~\cite{rfc8446, rfc6698}, RPKI validators already fetch entire repositories and must process Manifests and CRLs each cycle.
    This provides a natural metadata channel to carry shared verification material once, rather than duplicating it inside every object.

    \item \textit{Repository-centric distribution.}
    RPKI artifacts are fetched from publication points over HTTPS~\cite{rfc8182}, not carried in a handshake.
    Small metadata overheads that enable aggregation and batching are more acceptable here than in handshake-bounded settings~\cite{kwiatkowski-2019-pqc}.

    \item \textit{Bulk validation economics.}
    RPs validate at repository scale each cycle.
    Centralizing verification material lets validators reconstruct shared structure bottom-up and reuse intermediate nodes across many leaves, avoiding repeated per-object path processing~\cite{sikeridis-2020-overhead}.
    At the same time, \ours preserves per-object hooks where needed (\eg for non-public RSCs).

    \item \textit{Existing delegation hierarchy.}
    The RIR$\rightarrow$LIR structure provides clear administrative boundaries and naturally supports multi-level aggregation.
\end{packed_itemize}

\subsection{\ours: adapting MTL to RPKI}
\label{sec:substrate}

\subsubsection{Merkle Tree Ladder (MTL)}
\begin{figure}[t!]
    \centering
    \epsfig{figure=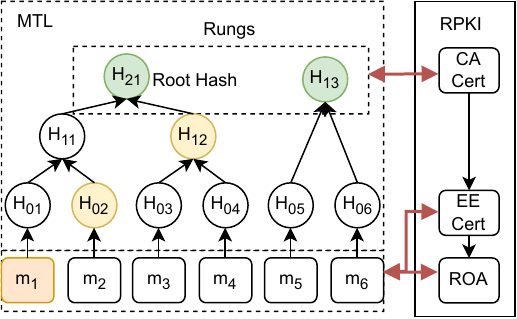,width=0.43\textwidth}
    \caption{Abstracting RPKI objects to Merkle Tree Ladder (MTL): a message $m_1$ is signed at a leaf; the verifier uses the authentication path (sibling hashes labeled yellow) to recompute the Merkle tree root.}
    \label{fig:sphincs}
\end{figure}

MTL~\cite{harvey-2023-mtl} amortizes authentication across many objects.
Rather than attaching a full public-key signature to every object, MTL hashes many objects into Merkle trees and protects a compact ladder root with a single signature.
To verify an object, the verifier recomputes the root using the object's leaf value and its Merkle \textit{authentication path} (\autoref{fig:sphincs}); \eg for $m_1$ the verifier combines a leaf value (\eg $H_{01}$) with sibling hashes (\eg $H_{02}, H_{12}$) to derive the rung root and then the ladder root.

A single fixed-depth Merkle tree supports only a fixed number of leaves; once exhausted, the tree must be rebuilt.
MTL avoids global rebuilds by linking a sequence of bounded-depth trees (``rungs'').
As new leaves are added, smaller rungs recursively merge into larger ones, yielding an extensible structure; for example, $14$ leaves are distributed across rungs of sizes $8, 4$, and $2$, which merge into a single rung of size $16$ once the next power of two is reached.

\subsubsection{Where per-object authentication lives in \ours}
In typical MTL-style deployments that prioritize standalone, per-object validation (\eg DNSSEC-oriented designs~\cite{fregly-2025-slh-dsa-mtl-dnssec}, the leaf index $i$ and authentication path $\Pi_i$ are carried with each record/certificate.
In RPKI, that approach is a poor fit: repositories contain hundreds of thousands of objects, so attaching tens of hashes per object (\eg $\sim$640\,B with SHA-256) would bloat \emph{every} object and turn routine churn into widespread reissuance.

\ours instead exploits the fact that RPKI validation is {\em repository-centric} and {\em manifest-driven}: RPs \textit{must} fetch Manifests/CRLs every cycle and validate in bulk. 
Thus, \ours relocates $i$ and $\Pi_i$ \textit{out of certificates/objects and into the Manifest}.
This yields three benefits:

\begin{packed_enumerate}
  \item \textit{No re-sign storm.}
  Adding an object or appending a rung changes only the Manifest (a metadata object) and the small set of internal nodes on the affected ladder path---not every object's certificate.
	  \item \textit{Small, stable objects.} Per-object size no longer tracks ladder evolution: each object carries only the minimal leaf-level material needed to recompute its leaf commitment, while indices and authentication information are supplied via the Manifest (through list order and $H_{0i}$).

  \item \textit{Bulk verification wins.}
  Because RPs validate many objects per cycle, they can reconstruct subtrees bottom-up and reuse intermediate nodes across leaves, instead of re-walking one authentication path per object.
  At the same time, \ours preserves per-object semantics needed for special cases (\eg RSC validation/revocation).
\end{packed_enumerate}

Moving $i$ and $\Pi_i$ does \textit{not} weaken integrity.
The Manifest is itself authenticated (as a depth-0 rung); each Manifest entry binds the \textit{file-list order} (inducing $i$) to a \textit{leaf commitment} $H_{0i}$; the object recomputes the same $H_{0i}$ from its bytes/header; and the reconstructed ladder root must match the CA's ladder root and, transitively, the parent aggregate.
Any mix-and-match attempt (wrong file at position $i$, stale path material, or stale Manifest) is rejected.

\subsubsection{MTL layout in \ours}
We split the ladder into two rung types to match RPKI's repository contents and refresh behavior:
\begin{packed_enumerate}
  \item \textit{Object rungs (published routing objects).}
  All published CA objects (\eg \roas) are leaves in that CA's ladder; the CA's ladder root authenticates the full set.
  This matches RPKI's bulk-fetch/bulk-validate environment: RPs already download and validate \textit{all} such objects each cycle.
  \item \textit{Metadata rungs (frequently updated support objects).}
	  The Manifest and CRL are modeled as single-leaf, {\em depth-0 rungs} appended after object rungs.
		Because Manifests and CRLs are reissued far more frequently than routing objects, placing them inside object rungs would make routine refreshes propagate through Merkle roots and trigger widespread recomputation (and, transitively, re-authentication) even when no RPKI objects change.
		Modeling each as a depth-0 rung isolates this churn to a short ladder suffix: only the metadata leaf and its upward path change, while all object leaves, indices, and cached subtrees remain stable.
		
\end{packed_enumerate}

These choices enable:

\begin{packed_enumerate}
  \item \textit{Rung sizing without per-object churn.}
  Since $i$ and $H_{0i}$ live in the Manifest, rung sizes can be tuned for signing cost and update locality without inflating objects.
  Appending a rung touches only a small number of nodes plus the Manifest.
  \item \textit{Deletion and hidden leaves without reindexing.}
  Physical removal would reindex the ladder.
  \ours preserves indices by using Manifest placeholders:
  mark deletions as $\mathsf{D}$ while retaining $H_{0i}$, and record revocation in the CRL;
  mark non-public artifacts (\eg RSCs) as $\mathsf{H}$ while retaining $H_{0i}$ so the ladder remains reconstructible without revealing content.
\end{packed_enumerate}

\subsubsection{Parent--child aggregation in the delegation hierarchy.}
Most CAs operate in \textit{hosted} mode, where a parent registry (\eg an RIR) aggregates child ladder roots and authenticates the aggregate with a single trust-anchor signature preconfigured at RPs.
In \textit{delegated} mode, a child maintains its own top-level key; the parent includes that key as a leaf in the aggregate, decoupling updates across tiers with minimal overhead (one Falcon public key and signature is about 1.53~KB).

\begin{figure}[t!]
    \epsfig{figure=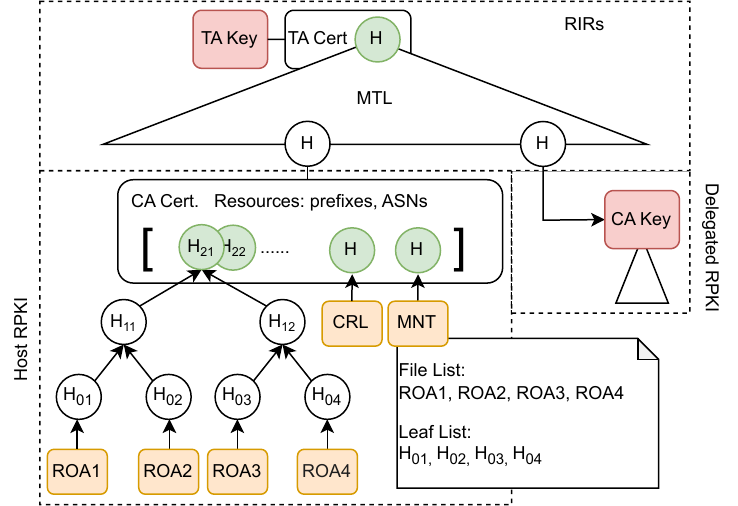,width=0.48\textwidth}
    \caption{\ours overview: a CA's objects (\eg \roas, ASPAs) are leaves of its ladder; the Manifest and CRL are appended as depth-0 rungs. At the registry level, a parent (\eg an RIR) aggregates child ladder roots and authenticates the aggregate with its trust anchor.}
    \label{fig:overview}
\end{figure}

\subsection{\ours: CAs and Publication Points}
\label{sec:structure-ours}

RPKI's hosted/delegated CA model is central to deployability.
\ours uses the same ladder and manifest abstraction to support both modes with the same RP verification logic, while enabling (i) append-only updates with stable indices, (ii) small and stable objects, and (iii) aggregation that does not couple parent and child update schedules (\autoref{fig:overview}).

\subsubsection{RPKI CA delegation modes}
\leavevmode\vspace{0.1cm}\par\noindent % Forces a break and adds controlled space

\ourparagraph{Hosted mode (RIR-managed CA).}
In hosted RPKI, the RIR operates the LIR's CA and repository.
Operationally:
\begin{packed_enumerate}
  \item \textit{Issue and retire objects.}
  The operator uses the RIR portal/API to create or withdraw \roas/\aspas.
  In current RPKI, each published object's SHA-256 hash appears in the Manifest; in \ours we reuse this value as the leaf commitment $H_{0i}$ in dual-deployment mode, providing both RSA and PQ integrity without duplicating object contents.
  In PQC-only mode, a CA can omit RSA certificates/signatures and apply \ours directly to the existing RPKI payload bytes.
  \item \textit{Update the ladder.}
  The RIR appends new leaves, recomputes only affected internal nodes, and derives the new ladder root for the hosted CA.
  \item \textit{Reissue metadata.}
  The Manifest (and, if needed, the CRL) is produced as a depth-0 rung carrying the ordered file list and leaf commitments $H_{0i}$ (\S\ref{sec:manifest}).
  \item \textit{Publish.}
  The PP performs an atomic update so RPs observe a consistent tuple: \{objects, Manifest, CRL\}.
  \item \textit{Aggregate at the registry.}
  The RIR updates the hosted child's entry in its registry aggregate to the child's new ladder root and signs the aggregate root with its trust anchor using a PQ-safe algorithm (\eg Falcon).
\end{packed_enumerate}

\ourparagraph{Delegated mode (LIR-managed CA).}
In delegated RPKI, the LIR runs its own CA and repository.
Steps (1)--(4) above are identical, but the parent inclusion differs:
\begin{packed_enumerate}
  \item[(e)] \textit{Parent inclusion.}
  The child signs and publishes its \textit{top-level ladder root}.
  The parent includes the child key as a leaf in the registry aggregate, so the child can update on its own cadence; the parent aggregate changes only when the child's \textit{key} rolls.
  This adds minimal bytes (about 1.53\,KB per delegated child; $\sim$97.7\,KB total given 59 delegated CAs as of Aug.~2025).
\end{packed_enumerate}

Across both modes, validators get the same assurances:
(i) the Manifest is an authenticated depth-0 rung and is verified first, so each cycle operates over a consistent object set;
(ii) leaf positions (indices $i$) persist across adds/deletes via placeholders ($\mathsf{D}$, $\mathsf{H}$), keeping object bytes and commitments $H_{0i}$ stable;
and (iii) updates are localized: only a short ladder suffix (and the Manifest/CRL) changes, while the parent aggregate continues to authenticate included children without requiring coordinated update timing.

\subsubsection{RPKI Objects Management}
\leavevmode\vspace{0.1cm}\par\noindent % Forces a break and adds controlled space
\ourparagraph{Manifest.} \label{sec:manifest}
\ours \textit{extends} the existing Manifest so that RPs can reconstruct and verify a CA's ladder in bulk without embedding per-object paths in certificates.
The extension is additive: legacy fields remain, and \ours relies on two Manifest-carried elements:
\begin{packed_itemize}
  \item \textit{Ordered file list.} The list order matches ladder leaf order and induces the index $i$; entries may be placeholders for deleted or non-public objects.
  \item \textit{Leaf commitments.} Each object's existing SHA-256 hash in the Manifest is reused as the leaf commitment $H_{0i}$ for position $i$ during ladder reconstruction.
\end{packed_itemize}

Placing $i$ and $H_{0i}$ in the Manifest prevents re-sign storms: adding objects or rotating rungs updates only this metadata rung and a short ladder suffix.
Moreover, a root-only metadata object is brittle under partial loss; carrying per-leaf commitments (and optionally internal nodes) makes recovery robust and lets validators fetch only missing pieces.

\ourparagraph{Non-public objects.}
Some CA-issued artifacts (\eg RSCs) must not be published.
\ours supports this via \textit{hidden} entries that preserve integrity without revealing filenames or content:
\begin{packed_itemize}
  \item In the file list, use placeholder $\mathsf{H}$ instead of a filename.
  \item In the leaf-commitment list, include the corresponding $H_{0i}$ so the ladder remains reconstructible and public objects remain verifiable.
\end{packed_itemize}
This keeps sensitive files off public repositories while preserving index stability and revocation hooks via the CRL.

\ourparagraph{Signing new objects.}
Publishing new objects follows an append-only pattern:
\begin{packed_enumerate}
  \item Append new leaves and compute only newly affected internal nodes (existing nodes remain unchanged).
  \item Update the Manifest's ordered file list and per-leaf $H_{0i}$ entries; reissue the Manifest rung.
  \item (Hosted) Refresh the child entry in the registry aggregate immediately; (Delegated) the child updates independently, and the parent aggregate changes only on child key rollover.
\end{packed_enumerate}
New \roas/ASPAs do not force re-signing unrelated objects; bytes and work scale with the number of new leaves.

\begin{figure}[t!]
    \centering
    \epsfig{figure=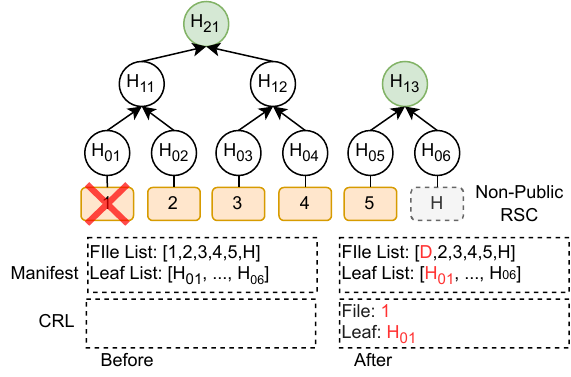,width=0.45\textwidth}
    \caption{Deletion without reindexing: the Manifest marks $\mathsf{D}$ while retaining $H_{0i}$, and the EE certificate is revoked via the CRL.}
    \label{fig:deletion}
\end{figure}

\ourparagraph{Deleting and revocation.}
Physical removal would reindex the ladder.
\ours preserves indices and expresses removals via metadata (\autoref{fig:deletion}):
\begin{packed_itemize}
  \item \textit{Delete the file.} Remove the file from the repository so it is no longer retrievable.
  \item \textit{Mark in the Manifest.} Replace the filename with placeholder $\mathsf{D}$ while retaining $H_{0i}$ so reconstruction and indices are unchanged.
  \item \textit{Revoke.} Add the corresponding EE certificate to the CRL.
\end{packed_itemize}

Retaining $H_{0i}$ for deleted entries does not enlarge the replay surface relative to today's RPKI because:
(i) the current, authenticated Manifest marks the entry as $\mathsf{D}$;
(ii) any stale file fails validation against the current Manifest's ordered list and its binding to $H_{0i}$;
(iii) the CRL independently signals revocation; and
(iv) repository transport is protected, preventing silent rollback of the Manifest.

\subsection{\ours: Relying Parties}
\label{sec:ours-validators}
We now switch to the RP perspective.
Each validation cycle must (i) detect whether a CA (or a registry aggregate) changed, (ii) fetch only what is needed, and (iii) validate efficiently.
\ours enables this with (a) compact ladder-root commitments at each CA and across aggregates, and (b) an authenticated Manifest that carries per-leaf commitments $H_{0i}$ for fine-grained change detection.

\subsubsection{Top-down ladder-guided synchronization}
\label{sec:sync}
RRDP deltas are efficient with warm caches, but after divergence (\eg restarts or missed serials) validators fall back to full snapshots even if most files are unchanged.
Large repositories therefore retain long delta histories to keep validators warm (\eg APNIC serves 500 deltas in its notification file).

\ours avoids dependence on delta retention by using authenticated ladder roots and the Manifest to localize changes:

\begin{packed_enumerate}
  \item \textit{Root and Manifest check (top-down).}
  Fetch the current ladder root for each registry aggregate (and the CAs it contains) and compare it to the cached root; if it matches, stop.
  If it differs, fetch and verify the corresponding Manifest.
  Because the aggregate commits to each child CA's ladder root, the RP can first identify which children changed by descending the aggregate, avoiding upfront Manifest fetches for unchanged CAs.

  \item \textit{Diff localization.}
  Compare the Manifest's ordered file list and leaf commitments $H_{0i}$ against the local cache to identify changed, deleted ($\mathsf{D}$), or hidden ($\mathsf{H}$) entries and affected subtrees.
  For a registry aggregate, each child CA appears as a leaf; this pinpoints which children changed even when there are tens of thousands.
  \item \textit{Targeted fetch.}
  Download only missing/changed files; remove entries marked $\mathsf{D}$; $\mathsf{H}$ requires no fetch.
  \item \textit{Rebuild \& cache.}
  Recompute the necessary leaf commitments and internal nodes bottom-up to the root; update the cache for the next cycle.
\end{packed_enumerate}

From an empty cache, an RP fetches the Manifest and the listed files once and reconstructs the ladder.
Afterwards, each cycle needs only the new Manifest plus changed files.
After restarts (RP or PP), the Manifest still provides a complete authenticated view, so catch-up does not depend on long RRDP delta retention.

\subsubsection{Bottom-up bulk verification}
\label{sec:bulk-verify}
Na\"ive MTL verification walks one authentication path per object and repeatedly rehashes shared internal nodes.
\ours verifies in bulk:
\begin{packed_itemize}
  \item \textit{Bottom-up reconstruction.}
    Compute leaf commitments for changed or newly fetched objects, then build internal nodes once per level, reusing them across leaves.
  The internal nodes reconstructed during synchronization can be reused for the final verification check.

  \item \textit{Manifest-driven indices.}
  The Manifest order induces indices $i$; each object is checked against its $H_{0i}$ and the tree is reconstructed to the root.
  \item \textit{Final root check.}
  The reconstructed root must match the CA's ladder root (and, for aggregates, the parent's signed aggregate root).
\end{packed_itemize}

\ourparagraph{Complexity and robustness.}
Per-cycle work scales with the number of changed leaves (plus affected subtree hashing), not total object count.
Because the Manifest carries the ordered file list and $H_{0i}$, missing files are detectable and localizable; $\mathsf{H}$ preserves reconstructability without disclosure; $\mathsf{D}$ is enforced via the CRL.
Any root mismatch or Manifest verification failure aborts the cycle, matching current RPKI safety semantics.

\subsection{Backward-compatible Deployment}
\label{sec:migration-rsa}

\ours preserves current \roa encodings and object payloads, so legacy validators can continue validating RSA unchanged while upgraded validators verify \ours in parallel; (1) on the publisher side, objects remain byte-for-byte compatible: publishers add \ours authentication alongside existing RSA signatures and (2) the Manifest is extended (not replaced) to carry indices $i$ while retaining all legacy fields, enabling incremental RP upgrades with an RSA fallback.
Once PQ validation is ubiquitous, RSA can be deprecated and removed; \ours can authenticate the same RPKI payload without RSA material.

\ourparagraph{Composability with repository compression.}
Repository-shrinking proposals that prune fields, remove redundancy, or change encodings~\cite{schulmann-2025-pruning, ddoesburg-2025-pqc} are {\em orthogonal} to \ours.
Because the ladder authenticates whatever bytes a CA publishes, any byte savings translate directly into smaller leaves.
\ours also remains applicable as long as revocation semantics for non-public artifacts (\eg RSCs~\cite{rfc9323}) are preserved: placeholders ($\mathsf{H}$, $\mathsf{D}$) and per-leaf commitments $H_{0i}$ still provide stable indexing and end-to-end binding even if EE certificates are removed.
Finally, synchronization accelerators that compare object hashes (\eg an Erik-style layer~\cite{snijders-2025-erik}) can reuse the Manifest's $H_{0i}$ commitments: the ladder root serves as a per-CA anchor, while $H_{0i}$ enables fine-grained change detection and targeted fetch with no additional metadata.

%% file: evaluation.tex
\section{Evaluation}
\label{sec:evaluation}

We implement \ours and perform a comprehensive end-to-end evaluation using real-world RPKI repositories. 
Our evaluation is designed to demonstrate that \ours is not merely a cryptographic substitution, but a structural optimization that resolves the tension between post-quantum security and operational scalability. We specifically address three critical operational imperatives:

\begin{packed_enumerate}
    \item \textit{Scalability:} Can \ours maintain a manageable repository footprint during the necessary dual-stack (RSA+PQC) transition?
    \item \textit{Operational Agility:} Does our optimized multi-level MTL design---specifically the isolation of metadata in \textit{Depth-0 rungs}---support the high-frequency signing cycles required by modern CAs?
    \item \textit{Efficiency:} Can the manifest-driven workflow reduce synchronization latency and enable faster global convergence compared to the current RRDP protocol?
\end{packed_enumerate}
\subsection{Implementation}
\label{subsec:implementation}

To evaluate \ours, we developed a complete prototype comprising both a CA and an RP.

\subsubsection{CA and Publication Point}
The authoritative workflow is divided into two stages:
\begin{packed_itemize}
    \item \textit{Object generation.} \ours preserves the existing RPKI object formats. We use Krill~\cite{Krill} to generate standard object payloads. This ensures that the underlying business logic of RPKI remains untouched and backward compatible.
    We implement a C++ MTL-based signer with \ours modifications: per-object verification material is moved into the Manifest, and the frequently refreshed Manifest/CRL are modeled as \textit{depth-0 rungs}.
    For hosted operation, the RIR signs hosted child CAs bottom-up and aggregates their ladder roots in the parent MTL.

    \item \textit{Publication (PP).} Same with original RPKI, PP serves objects over HTTPS (using Nginx). 
    In PQ-only mode, RRDP notifications and deltas are not needed because validators compute differences on-the-fly using the Merkle Tree.
\end{packed_itemize}

\subsubsection{Relying Party (Validator)}
The validator prototype implements a ladder-guided workflow designed to minimize computational redundancy:
\begin{packed_itemize}
    \item \textit{Synchronization.} The RP maintains a local cache and fetches updates from PPs. We implement the top-down ladder-guided synchronization protocol based on the multi-level MTL structure.
    \item \textit{Validation.} We implement a bottom-up bulk verification engine in C++. Rather than verifying each object independently,
    our engine verifies objects against $H_{0i}$ and then reconstructs internal nodes level-by-level. This ensures each internal node is computed exactly once per cycle.
\end{packed_itemize}

\ourparagraph{Environment.}
Signing and validation benchmarks run on an Intel Xeon Silver 4210R @ 2.40\,GHz with 187\,GB RAM (Ubuntu 22.04.5 LTS).
Synchronization and end-to-end tests use AWS-hosted PPs and on-prem validators over 1\,Gbps links with RTT $<10$\,ms, so the network is not the bottleneck.

\subsection{Experimental methodology and datasets}
\label{subsec:methodology}

Our evaluation is multi-axis because a PQ transition in RPKI spans three
dimensions: (i) authentication structure (per-object signatures vs. Merkle
aggregation), (ii) deployment mode (PQC-only vs. dual with RSA), and (iii)
repository layout (integrity-preserving vs. compressed). 
We therefore evaluate four schemes under two deployment modes on both layouts
to separate cryptographic overhead from architectural effects and to quantify
the cost of dual deployment.

\ourparagraph{Authentication schemes.}
We consider:
\begin{packed_enumerate}
  \item \textit{RSA baseline (Original RPKI).}
  We use Krill as the CA/PP and \texttt{rpki-client} as the RP.
  Following prior measurements~\cite{sediqi-2025-rpki}, \texttt{rpki-client} outperforms Routinator~\cite{Routinator}, and we therefore treat it as our RP baseline.

  \item \textit{Multi-use PQ signatures (na\"ive PQ swap).}
  We evaluate FN-DSA (Falcon) and ML-DSA (Dilithium) by signing and verifying the \textit{same} repository objects using OpenSSL, without changing object payloads or repository organization.
  This baseline captures the practical ``cost of inaction": the bandwidth and verification overhead incurred when every object carries a multi-use PQ signature.

  \item \textit{Native MTL.}
  We implement MTL as originally specified~\cite{harvey-2023-mtl}, without the RPKI-specific changes introduced in \autoref{sec:substrate} 
  The ladder root is authenticated with Falcon due to its smaller public key and signature than ML-DSA.
  Because native MTL does not provide our Manifest-driven change-localization interface, we compare it on repository size and CA/RP signing/validation costs, but do not report RRDP-style synchronization or PP-to-router end-to-end time.

  \item \textit{\ours.}
  Our full design and prototype, including Manifest-carried leaf commitments $H_{0i}$ (and list-induced indices), depth-0 rungs for Manifest/CRL, and ladder-guided synchronization with bottom-up bulk verification.
  As with native MTL, we use Falcon to authenticate the ladder root to minimize public-key and signature overhead.
\end{packed_enumerate}

\ourparagraph{Repository layouts.}
Unless otherwise stated, results use the \textit{Original} (integrity-preserving) RPKI object layout.
We additionally report results over two compressed layouts (null scheme and iRPKI) to quantify how \ours composes with payload-reduction techniques.
These compressed layouts are \textit{not currently} drop-in replacements for today's operational RPKI: they remove or alter certificates/fields and are not accepted as-is by the working group.
However, they provide a useful stress test: any byte reductions translate directly into smaller leaves and less hashing and transfer work in \ours.

\ourparagraph{Deployment modes.}
A realistic evaluation must account for the transition period from classical to post-quantum cryptography. We evaluate two distinct deployment scenarios to reflect both the transition phase and the end-state:
\begin{packed_enumerate}
	\item \textit{Dual-deployment Mode:} In this transitional phase, CAs publish both classical RSA and post-quantum signatures for every object. Given that RSA-2048 remains secure against classical threats and will likely coexist with PQC for a prolonged period (similar to hybrid models in other PKI ecosystems such as \cite{GooglePQCThreatModel,cloudflare-2025-pqc}), this is the expected operational model for the foreseeable future. Therefore, this mode serves as the \textit{primary focus} of our evaluation regarding operational feasibility.

    \item \textit{PQC-only Mode:} A forward-looking scenario where CAs publish only post-quantum signatures. We use this mode to evaluate the pure efficiency of the architectures once the transition from RSA is complete.
\end{packed_enumerate}

\input{tables/size.tex}
\ourparagraph{Datasets.}
We curated two datasets because storage scalability and update agility require different stress tests.
\begin{packed_itemize}
    \item \textit{7-day Dataset (Steady State Analysis):} We captured the complete state of global RPKI every 20 minutes for 7 days (Aug 6-13, 2025). Across this period, we observe 64 publication points in total (5 RIRs + 59 delegated CAs) containing between 465,646 and 470,303 objects. On the last day, the object mix included 319,186 ROAs, 49,263 Manifests, 49,262 CRLs, and 47,739 certificates, yielding an average footprint of 853.2 MB across all PPs.
    \item \textit{1-day High-Frequency Dataset (Agility Stress Test):} Standard RPKI datasets do not capture minute-level dynamics. To evaluate the agility of our signing architecture, we collected a dataset with a 1-minute interval (1,440 snapshots). This dataset is used exclusively to evaluate \textit{Per-minute Update Signing Time}, ensuring our design can withstand high-frequency policy changes without latency spikes.
\end{packed_itemize}

\subsection{Results}
\subsubsection{Repository Scalability}
\label{sec:res-repo-size}

The most immediate threat posed by the PQC transition is repository bloat; we analyze the repository footprint using the 7-day steady-state dataset.

\ourparagraph{Feasibility of Dual-Stack Deployment.}
\begin{figure}[t!]
    \centering
    \epsfig{figure=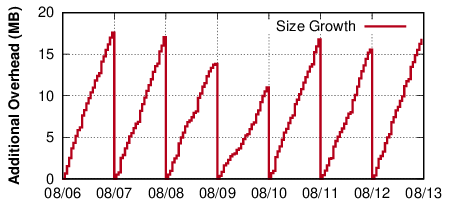,width=0.48\textwidth}
    \caption{Cumulative extra bytes from revoked objects over one week. The sawtooth pattern shows bounded overhead that resets daily.}
    \label{fig:addition-size}
\end{figure}
A smooth transition requires a ``Dual-Stack" period where CAs publish both RSA and PQC signatures. As shown in \autoref{tab:rpki-size}, a na\"ive application of ML-DSA causes the repository to balloon from 853.2 MB to 3,577.3 MB, which is a staggering increase of +319.3\%. Such growth would require significant infrastructure upgrades globally.
In stark contrast, \ours demonstrates exceptional efficiency. In {Dual-deployment mode}, \ours averages 882.4 MB, adding only 29.2 MB (+3.4\% overhead) compared to the current RSA-based RPKI. This near-zero overhead is achieved because \ours reuses the existing SHA-256 object hashes as leaf commitments ($H_{0i}$) for the PQC ladder. This means the PQC security layer is effectively ``grafted" onto the existing RPKI payloads without duplicating data, solving the backward compatibility challenge without the storage penalty.

\ourparagraph{Synergy with Compression Schemes.}
We also evaluated \ours in conjunction with proposed compression schemes (Null Scheme and iRPKI) to assess compatibility. In a dual-stack scenario, combining \ours with iRPKI reduces the repository size to just {159.1 MB}. This represents an {81.4\% reduction} from the current uncompressed RSA baseline (853.2 MB). This confirms that \ours is orthogonal to payload optimization; operators can deploy \ours for PQ-security and optionally apply compression to further minimize bandwidth, providing a flexible path forward for resource-constrained environments.

\ourparagraph{Architectural Efficiency vs. Native MTL.}
The advantage of \ours's design becomes even clearer in the PQC-only mode. \ours reduces the total footprint to 546.8 MB, which is 306.4 MB smaller (-35.9\%) than today's RSA RPKI.
Crucially, \ours is significantly smaller than the Native MTL baseline (836.1 MB).
This reduction stems from our architectural decision to eliminate redundant path data. Native MTL designs embed an authentication path (sibling hashes) into every single object. As the tree depth increases, this per-object overhead accumulates linearly. By consolidating these paths into the centralized Manifest, \ours eliminates this massive redundancy. Our data shows this choice alone yields a 34.6\% reduction in size compared to generic MTL.

\ourparagraph{Intra-day Growth Stability.}
\ours uses ``deletion placeholders" ($\mathsf{D}$) in the Manifest to preserve tree indices during the day, which simplifies signing but technically retains ``dead" references until the daily re-sign.
One might concern this leads to bloat. \autoref{fig:addition-size} analyzes this intra-day growth. We find the accumulated overhead peaks at only 17.5 MB (approximately 2.0\% of the total size) before being reset.
This confirms a favorable engineering trade-off: we accept a trivial 2\% temporary storage cost to gain the massive performance benefits of append-only signing, which avoids the computational cost of re-indexing the entire tree for every revocation.

\subsubsection{Operational Agility: Signing Time}
\label{sec:res-signing}

Agility is defined by the ability to process updates quickly.

\begin{figure}[t!]
    \centering
    \epsfig{figure=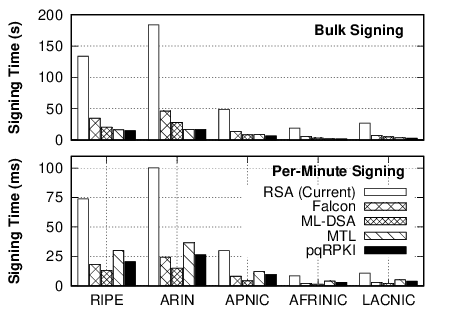,width=0.48\textwidth}
    \caption{Signing latency for bulk reissuance (top) and per-minute updates (bottom).}
    \label{fig:signing-time}
\end{figure}

\ourparagraph{Bulk Signing Performance by Registry.}
\autoref{fig:signing-time} (Top) presents a breakdown of bulk signing times across the five RIRs.
\ours consistently outperforms the current RSA baseline by an order of magnitude.
For ARIN (the largest repository), the signing time drops from 183.8 seconds (RSA) to just 16.8 seconds with \ours, a \textit{90.9\% reduction}.
Similarly, RIPE signing time decreases from 133.6 seconds to 14.9 seconds (-88.8\%).
Even compared to Native MTL (\eg 16.1 s for RIPE), \ours is faster;
this drastic reduction is due to the shift from expensive modular exponentiation (required for RSA signatures on every object) to efficient hash-based aggregation. Even compared to ML-DSA (28.0 s for ARIN), \ours is faster as it aggregates thousands of objects under a single ladder root signature.

\ourparagraph{Incremental Updates and the Depth-0 Advantage.}
The more critical metric for operations is the {per-minute update} time. As shown in \autoref{fig:signing-time} (Bottom), \ours processes updates in just 112.3 ms on average.
While pure ML-DSA signing is naturally faster (63.3 ms) due to efficient lattice arithmetic, it comes at the unacceptable cost of 3.6 GB storage bloat.
\ours represents the optimal balance: it is 3.7$\times$ faster than the current RSA system (412.1 ms) while maintaining minimal storage.
The slight overhead compared to ML-DSA reveals the cost of maintaining a structured repository: every update requires regenerating the Manifest and re-hashing the path. However, our {Depth-0 rung} design ensures this cost remains constant ($O(1)$) regardless of repository size. Updating the Manifest (a Depth-0 node) does not require re-hashing the stable object tree, effectively decoupling the cost of metadata churn from the size of the repository.

\subsubsection{Efficiency: Validation and Synchronization}
\label{sec:res-validation}

\input{tables/validation.tex}

\autoref{tab:validation} presents the total time to validate the global repository using the 7-day dataset. \ours completes this in {102.7 seconds}, which is {51.8\% faster} than the Original RSA time (213.2 s) and roughly 4$\times$ faster than ML-DSA (404.6 s).
Notably, \ours outperforms {Native MTL} (158.1 s) by approximately 35.0\%.
This gap highlights the efficiency of our \textit{bottom-up reconstruction}. A Native MTL verifier works per-object: for $N$ objects, it re-hashes the path to the root $N$ times ($O(N \log N)$), performing redundant computations on shared internal nodes. The \ours verifier, guided by the Manifest, rebuilds the tree level-by-level, hashing each internal node exactly once ($O(N)$) regardless of how many leaves sit below it.

\subsubsection{Synchronization and PP-to-router convergence}
We benchmark synchronization over the full 7-day dataset.
For these measurements, we use PQC-only mode: in dual-stack deployments, RSA signatures increase the bytes transferred, but they do not change the underlying synchronization mechanism or its latency behavior.
Thus, PQC-only mode provides the cleanest comparison between \ours's Manifest-guided design and standard RRDP.

\ourparagraph{Synchronization latency.}
Current RPKI relies on RRDP, which distributes updates via XML notification and delta files; at scale this adds XML parsing overhead and requires maintaining long delta histories.
\ours instead uses Manifest-guided synchronization.

\begin{packed_itemize}
    \item \textit{Cold sync (empty cache):}
    RSA takes 11.4~s; a na\"ive PQ swap increases this due to object inflation (Falcon 18.2~s, ML-DSA 25.6~s).
    \ours completes cold sync in 6.1~s, 1.9$\times$ faster than RSA (-46.5\%) and 4.2$\times$ faster than ML-DSA (-76.2\%), primarily because fewer bytes are transferred.

    \item \textit{Warm sync (incremental):}
    In steady state (about 1,201 changed objects or 2.9~MB per interval), RSA takes 4.1~s and \ours takes 2.8~s (1.5$\times$, -31.7\%); PQ baselines remain slower (Falcon 6.7~s, ML-DSA 9.4~s).
    Unlike RRDP's fetch-parse-apply chain (notification $\rightarrow$ delta URL $\rightarrow$ delta XML), \ours fetches ladder roots top-down and compares the Manifest's $H_{0i}$ list against the cache to localize additions, deletions ($\mathsf{D}$), and modifications without XML processing.
    This allows the RP to identify any updates immediately without XML processing overhead. 
\end{packed_itemize}

\ourparagraph{End-to-end propagation.}
We measure ``time-to-protection'': the time from publication at the PP to emitting VRPs for routers (synchronization plus bulk validation).
Over the 7-day corpus, the RSA baseline averages 226.9~s, while a na\"ive ML-DSA swap rises to 431.5~s.
\ours averages 118.3~s, a 1.9$\times$ reduction (-47.9\%) versus RSA and 3.6$\times$ faster than ML-DSA.
This suggests the ecosystem could move from today's 20--40 minute update interval to sub-2-minute operation, sharply reducing the window between ROA issuance/revocation and router enforcement.
This significantly tightens the ``window of vulnerability"---the time between ROA issuance and revocation and its enforcement at the router level---thereby transforming RPKI into a near real-time defense mechanism.

%% file: tables/size.tex
\begin{table}[t]
\centering
    \fontsize{8}{10}\selectfont
\caption{RPKI repository size under different deployments (7-day average, MB).}
\begin{tabular}{ll|rrr}
\multicolumn{1}{c}{\multirow{3}{*}{Deployment}} &
\multicolumn{1}{c|}{\multirow{3}{*}{Scheme}} &
\multicolumn{3}{c}{Repository Size (MB)} \\
\cline{3-5}
 & & \multicolumn{1}{c}{Original} & \multicolumn{2}{c}{Compressed} \\
 & & \multicolumn{1}{c}{(Integrity)} & \multicolumn{1}{c}{(Null Scheme)} & \multicolumn{1}{c}{(iRPKI)} \\
\hline
Baseline & RSA
    & 853.2  & 685.8  & 142.6 \\
\hline
\multirow{4}{*}{PQ-only}
 & ML-DSA
    & 3241.6 & 2058.5 & 480.7 \\
 & Falcon
    & 1582.7 & 1108.4 & 229.4 \\
 & Native MTL
    & 836.1  & 645.6  & 130.7 \\
 & \ours
    & \textbf{546.8}  & \textbf{412.0}  & \textbf{83.9} \\
\hline
\multirow{4}{*}{\shortstack[l]{Dual-Stack\\w/ RSA}}
 & ML-DSA
    & 3577.3 & 2354.8 & 533.0 \\
 & Falcon
    & 1918.4 & 1404.7 & 260.8 \\
 & Native MTL
    & 1171.6 & 937.9  & 205.9 \\
 & \ours
    & \textbf{882.4}  & \textbf{704.3}  & \textbf{159.1} \\
\end{tabular}

\label{tab:rpki-size}
\end{table}

%% file: tables/validation.tex
\begin{table}[t]
	\centering
	\fontsize{8}{10}\selectfont
	\caption{Avg. time to validate \textit{all} RPKI objects per cycle.}
\begin{tabular}{c||cccc|c}
RPKI Objects & RSA   & Falcon & ML-DSA & MTL & \ours\\ \hline
%& \multicolumn{4}{c}{Average} \\\hline
Original & 213.2 & 303.9  & 404.6  & 158.1 & 102.7  \\ \hline
Compressed & & & & \\ \hline
Null Scheme            & 142.6 & 330.7  & 451.2 & 103.6  & 64.9  \\ 
iRPKI            & 31.5 & 48.0 &  59.3 & 24.1 & 19.7  \\
%& \multicolumn{4}{c}{Maximum} \\\hline
%Current ROA             & 235.6 & 339.5  & 451.3  & 171.2  \\
%100\% ROA           & 416.1 & 582.4  & 768.9  & 241.6  \\ \hline
\end{tabular}
%\caption{Average and maximum time (s) to validate \textit{all} RPKI objects per cycle (20 min cadence). \weitong{Maybe we can just keep the average?}}

\label{tab:validation}
\end{table}

%% file: discussion.tex
\section{Discussion}
\label{sec:discussion}
\ourparagraph{Consistency and transparency.}
Operators often run validators at multiple sites to detect discrepancies caused by PP misconfiguration, network faults, or active interference.
Today, most checks compare VRPs across sites (\eg rtrmon~\cite{StryRTR}, ByzRP~\cite{friess-2024-byzantine}), which forces each site to download and validate the full corpus before any comparison.
This is costly, slow, and hard to localize: when VRPs differ, operators still must pinpoint which CA, PP, or file diverged.
Exchanging raw objects is infeasible at scale, and even exchanging all Manifests is heavy (more than 49,263 files today).

\ours enables lightweight checks at the authentication layer.
Sites first exchange per-CA ladder roots (or registry aggregate roots), which are compact hashes; matching roots imply identical authenticated views with negligible bandwidth.
If roots differ, sites exchange only the Merkle nodes needed to bisect to the mismatch, isolating the divergent subtree and leaf positions.
Because leaves are bound to the Manifest's ordered file list and leaf commitments, mismatches map directly to concrete repository artifacts rather than opaque VRP diffs, improving time-to-diagnosis and supporting transparency services (\eg RPKIview~\cite{RPKIviews}) without redistributing full repositories.

\ourparagraph{Availability and stalling on the PP-to-RP path.}
Attackers can exploit the PP-to-RP interaction using deep delegation chains, slow or non-responsive publication, and related tactics that stretch RP timeouts and induce staleness~\cite{hlavacek-2022-stalloris, hlavacek-2023-beyond, gilad-2017-deployment, mirdita-2025-cure}.
Timeout tuning is a brittle defense: short timeouts can drop legitimate updates, while long ones can be abused to stall validators.

\ours reduces per-PP work and bytes via Manifest-guided synchronization (\autoref{sec:sync}), which helps validators make progress faster and adopt tighter per-PP timeouts in practice.
Core trust semantics {\em remain unchanged}: PPs still serve over HTTPS, and certificate chains, validity periods, and CRL checks follow today's RPKI rules.

\ourparagraph{Bounded growth from deletions and delegations.}
To avoid reindexing churn, \ours preserves leaf indices within each signing epoch.
When an object is removed, the Manifest marks its entry as deleted ($\mathsf{D}$) but retains the corresponding leaf commitment $H_{0i}$ until the next routine daily rebuild.
A misbehaving client could try to inflate the Manifest by rapidly issuing and revoking objects, most plausibly in hosted mode.
This is manageable without protocol changes: CAs can rate-limit update bursts and cap per-CA revocations (or total $\mathsf{D}$ entries) per rebuild window.
The daily rebuild then drops stale entries and keeps growth bounded (as quantified in \autoref{fig:addition-size}).
We accept this small, temporary overhead to enable append-only updates and deletion without reindexing, which directly improves signing and validation performance.

Delegated mode adds one PQ public key and signature per delegated child CA; since delegated CAs are rare in today's deployment, the footprint impact is small and can be bounded by operational policy (for example, capping delegated children per parent CA).

%% file: conclusion.tex
\section{Conclusion} \label{sec:conclusion}

We present \ours, a post-quantum RPKI framework that combines a multi-level Merkle Tree Ladder (MTL) with today's RPKI objects and relocates per-object verification material from certificates into the Manifest.
\ours introduces a ladder-guided synchronization and bulk-verification workflow, models the frequently refreshed Manifest/CRL as depth-0 rungs, and preserves \roa payloads while supporting both hosted and delegated operation and a phased migration that can coexist with RSA.
In PQC-only mode, \ours reduces average repository size by 65.5\%/83.1\% compared to directly applying FN-DSA/ML-DSA, and enables sub-2-minute operation with full-repository validation each cycle.

Looking ahead, we see clear paths to deployment: tuning per-CA parameters, integrating with sync layers (\eg Erik) and size-reduction encodings, and expanding operational trials to stress consistency and failure modes across diverse publication points.

%% file: paper.bbl
\newcommand{\etalchar}[1]{$^{#1}$}